\documentclass[prb,aps,nobalancelastpage,showpacs,amssymb,twocolumn,citeautoscript]{revtex4}    
\usepackage{mathrsfs}
\usepackage[figuresright]{rotating}
\usepackage{amsmath}
\usepackage{amssymb}
\usepackage{graphicx}
\usepackage{color}
\usepackage{dcolumn}
\usepackage{bm}
\makeatletter

\newcommand{\Rmnum}[1]{\expandafter\@slowromancap\romannumeral #1@}
\makeatother

\begin{document}
\preprint{PRB(R)/Y. K. Luo et al.}

\title{Li$_2$RhO$_3$: A spin-glassy relativistic Mott insulator}

\author{Yongkang Luo$^{1}$, Chao Cao$^{2}$, Bingqi Si$^1$, Yuke Li$^{2}$, Jinke Bao$^{1}$, Hanjie Guo$^{1}$, Xiaojun Yang$^{1}$, Chenyi Shen$^{1}$, Chunmu Feng$^{1}$, Jianhui Dai$^{2}$, Guanghan Cao$^{1}$, and Zhu-an Xu$^{1}$\footnote[1]{Electronic address: zhuan@zju.edu.cn}}

\address{$^1$Department of Physics and State Key Laboratory of Silicon Materials, Zhejiang University,
Hangzhou 310027, China,}
\address{$^2$Department of Physics, Hangzhou Normal University,
Hangzhou 310036, China.}

\date{\today}

\begin{abstract}

Motivated by the rich interplays among electronic correlation, spin-orbit coupling (SOC), crystal-field splitting, and geometric frustrations in the honeycomb-like lattice, we systematically investigated the electronic and magnetic properties of Li$_2$RhO$_3$. The material is semiconducting with a narrow band gap of $\Delta\sim$78 meV, and its temperature dependence of resistivity conforms to 3D variable range hopping mechanism. No long-range magnetic ordering was found down to 0.5 K, due to the geometric frustrations. Instead, single atomic spin-glass behavior below the spin-freezing temperature ($\sim$6 K) was observed and its spin dynamics obeys the universal critical slowing down scaling law. First-principles calculation suggested it to be a relativistic Mott insulator mediated by both electronic correlation and SOC. With moderate strength of electronic correlation and SOC, our results shed new light to the research of Heisenberg-Kitaev model in realistic materials.

\end{abstract}


\maketitle


Ternary transition metal oxides set up a fascinating platform for investigating correlated electronic systems. Depending on the particular transition metal element and crystalline structure, features like high-temperature superconductivity (SC) in doped spin-1/2 antiferromagnetic (AFM) Mott insulator La$_2$CuO$_4$\cite{Bednorz-La2CuO4}, giant negative magnetoresistance in La$_{2/3}$Ba$_{1/3}$MnO$_3$ films\cite{Helmolt-LaMnO3}, odd parity SC in
Sr$_2$RuO$_4$\cite{Maeno-Sr2RuO4,Nelson-Sr2RuO4}, SC in water intercalated Na$_x$CoO$_2$$\cdot$ $y$H$_2$O\cite{Takada-NaxCoO2}, and field induced metamagnetic transition and quantum criticality in Sr$_3$Ru$_2$O$_7$\cite{Perry-Sr3Ru2O7} have been observed. Electronic correlation is expected to be strongest in 3$d$
transition metals, represented by a small $d$ orbital radius and a large Coulomb repulsion $U$. It weakens as one goes from 3$d$ to 4$d$ and 5$d$ transition metals due to the spatial extension of $d$ orbits. However, the relativistic spin-orbit coupling (SOC) which increases with atomic number follows the opposite trend. The recently discovered exotic non-metal behaviors in those heavy transition metal oxides\cite{Yanagishima-R2Ir2O7,Erickson-BaNaOsO,Gegenwart-Na2IrO3,Kim-Sr2IrO4} remind us of the importance of SOC in these materials. One representative example is Sr$_2$IrO$_4$\cite{Kim-Sr2IrO4}, which was confirmed to be a novel Mott insulator mediated by strong SOC even though the electronic correlation is relatively weak, while its structural analog Sr$_2$RhO$_4$\cite{Perry-Sr2RhO4} shows normal Fermi-liquid metallicity.

\begin{figure}[htbp]
\includegraphics[width=1.0\columnwidth]{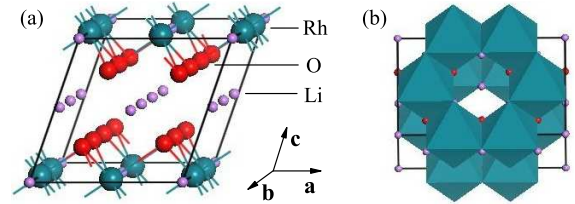}
\caption{(Color online)\label{Fig.1} (a), The crystalline structure of Li$_2$RhO$_3$, which is stacked by alternating Li and LiRh$_2$O$_6$ layers. (b), Within the LiRh$_2$O$_6$ layer, the RhO$_6$ octahedra form the honeycomb-like lattice. }
\end{figure}

The general formula Li$_2$$M$O$_3$ ($M=$ transition metal) actually describes two types of crystalline structures: Li$_2$MnO$_3$-type (C2/$m$, No.~12)\cite{Jansen-Li2MnO3} and Li$_2$SnO$_3$-type (C2/$c$, No.~15)\cite{Lang-Li2SnO3}. In both crystalline structures, the layers of $M$O$_6$ octahedral interstices are alternately filled either with Li$^{+}$ only, or with 1/3 Li$^{+}$ and 2/3 $M^{4+}$, as depicted in Fig.~\ref{Fig.1}(a), whereas the $M$O$_6$ octahedra with $M$ in the center form edge-sharing honeycomb-like networks\cite{VTodorova-Li2RhO3}[Fig.~\ref{Fig.1}(b)]. Tiny difference between these two crystalline structures resides in the stacking of Li-$M$ layers along $c$-axis: in the case of Li$_2$SnO$_3$-type, the Sn$^{4+}$ hexagonal networks in adjacent
layers are displaced by (0, $\pm$ 1/6, 1/2) in lattice coordinates, while in Li$_2$MnO$_3$-type, they are displaced by (0, 1/2, 1/2)\cite{Kobayashi-Li2RuO3}. The formation of honeycomb-like $M$O$_6$ octahedral network makes Li$_2$$M$O$_3$ a suitable candidate for investigating the interplays among electronic correlation, SOC, crystal-field splitting and geometric frustrations.

\begin{figure}[htbp]
\includegraphics[width=8cm]{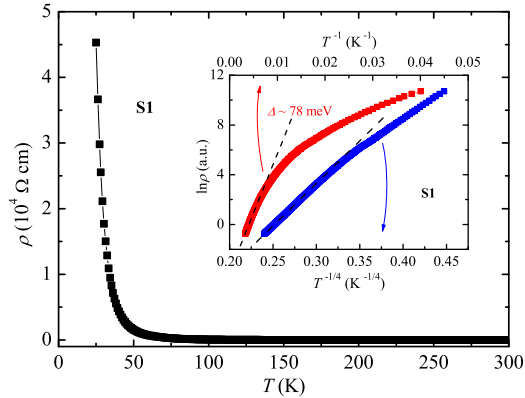}
\caption{(Color online)\label{Fig.2} Electrical resistivity of Li$_2$RhO$_3$. The inset shows resistivity in the Arrhenius plot and 3D-VRH plot. The dashed lines are guides to eyes. Thermal activating band $\Delta\sim$ 78 meV is estimated in the Arrhenius plot.}
\end{figure}

Li$_2$RhO$_3$ and Li$_2$IrO$_3$ crystalize in Li$_2$MnO$_3$- and Li$_2$SnO$_3$-type structures, respectively. Previously, they were studied for potential applications as Li-ion battery cathode materials\cite{VTodorova-Li2RhO3,Kobayashi-Li2IrO3,Malley-Li2IrO3}. More underlying physical properties still need to be explored. Herein, we systematically studied the electronic and magnetic properties of Li$_2$RhO$_3$. Our results point out that Li$_2$RhO$_3$ is likely to be a spin-glassy Mott insulator with a narrow thermal activating gap $\Delta\sim$78 meV, while the spin-freezing temperature is sample dependent ranging from 5 to 7
K. No long range magnetic ordering can be captured in this frustrated system down to 0.5 K. Our experimental results were understood by first principle calculations which confirm the important roles played by both electronic correlation and SOC. The calculation also pointed out that Li$_2$RhO$_3$ is on the boundary of antiferromagnetically-ferromagnetically correlated ground state, which might interpret the spin-glassy behavior observed experimentally.

\begin{figure*}[htbp]
\includegraphics[width=18cm]{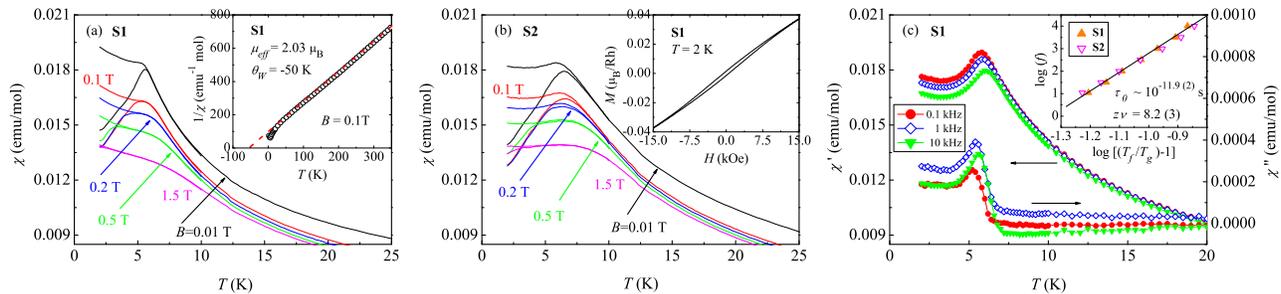}
\caption{(Color online)\label{Fig.3} Temperature dependent magnetic susceptibility of Li$_2$RhO$_3$. (a-b), $\chi(T)$ of \textbf{S1} and \textbf{S2} measured under various fields, respectively, in both ZFC and FC processes. Inset of (a) displays Curie-Weiss fit of $1/\chi(T)$ in the high $T$ region, while inset of (b) shows hysteretic loop in $M(H)$ below $T_g$. (c), ac susceptibility measurement of \textbf{S1}. Inset of (c): scaling plot of $\log(f)$ vs. $\log[(T_f/T_g)-1]$ for \textbf{S1} (solid) and \textbf{S2} (open), with the best fitted parameters $\tau_0\sim 10^{-11.9(2)}$s, $z\nu=$8.2 (3), $T_{g1}=$5.30 K, and $T_{g2}=$5.97 K. The solid line is a guide to eyes of this fitting. }
\end{figure*}


Poly-crystalline sample of Li$_2$RhO$_3$ was grown by solid state reaction method
as mentioned elsewhere\cite{VTodorova-Li2RhO3}. The sample quality of Li$_2$RhO$_3$ was checked by X-ray diffraction (XRD), performed on a PANalytical X-ray diffractometer (Empyrean Series 2) with Cu-K$_{\alpha 1}$ radiation at room temperature. Lattice parameters were derived by Rietveld refinement on the RIETAN-RF
programme\cite{Izumi-Rietan_RF}. Electrical resistivity and specific heat were measured on a Quantum Design physical property measurement system (PPMS-9). The dc magnetization measurement was carried out on a Quantum Design magnetic property measurement system (MPMS-5) employing both zero-field-cooling (ZFC) and field-cooling (FC) protocols. The ac magnetic susceptibility was measured on PPMS-9 with various frequencies ranging from 10 Hz to 10 kHz.

\begin{figure}[htbp]
\includegraphics[width=1.0\columnwidth]{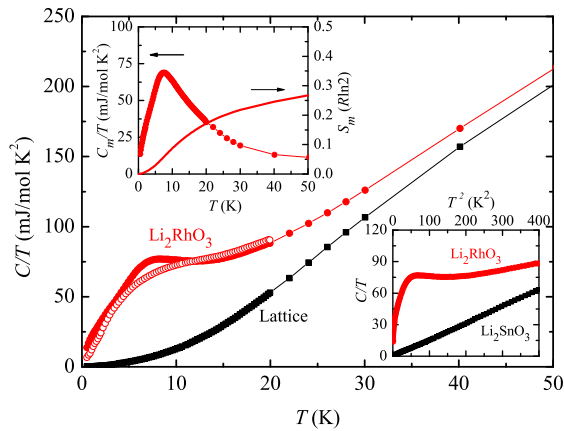}
\caption{(Color online)\label{Fig.4} Main frame: specific heat divided by $T$ of Li$_2$RhO$_3$ (red), compared with the lattice contribution (black) which was derived by correcting its non-magnetic reference Li$_2$SnO$_3$. The solid (open) symbols represent data measured under $\mu_0H=$0 (9 T). Lower inset: $C/T$ of Li$_2$RhO$_3$ and Li$_2$SnO$_3$ plot in the $T^2$ scale. Upper inset: magnetic contribution to specific heat in Li$_2$RhO$_3$, also shown is the magnetic entropy gain $S_m$ as a function of $T$.}
\end{figure}


The XRD pattern (not shown) guarantees high purity of the samples, and all the peaks can be well indexed based on the C2/$m$ (No.~12) space group iso-structural to Li$_2$MnO$_3$. The Rietveld refinement yields $a=$5.1212(3)\AA, $b=$8.8469(4)\AA, $c=$5.1015(3)\AA, $\alpha$=$\gamma=$90\textordmasculine, and $\beta=$109.641(3)\textordmasculine, which are comparable with those in the previous literature\cite{VTodorova-Li2RhO3}. Detailed structural parameters are summarized in Table~\ref{table1}. The labels \textbf{S1} and \textbf{S2} represent the two Li$_2$RhO$_3$ samples annealed at 950$\,^{\circ}\mathrm{C}$ and 900$\,^{\circ}\mathrm{C}$, respectively. It should be pointed out that there is an antisite disorder between Li$^{+}$ ions and Rh$^{4+}$ ions. Such anti-site disorder is a common feature in Li$_2M$O$_3$ materials\cite{Kobayashi-Li2RuO3,Kobayashi-Li2IrO3}, and may has a double-side influence on the magnetism: on the one hand, the partial substitution of $M^{4+}$ by non-magnetic Li$^{+}$ ions breaks the long range magnetic coupling among $M^{4+}$ moments, while on the other hand, it also reduces the geometric frustrations on the $M^{4+}$ honeycomb lattice and thus stabilizes the magnetic structure.

\begin{figure*}[htbp]
\includegraphics[width=18cm]{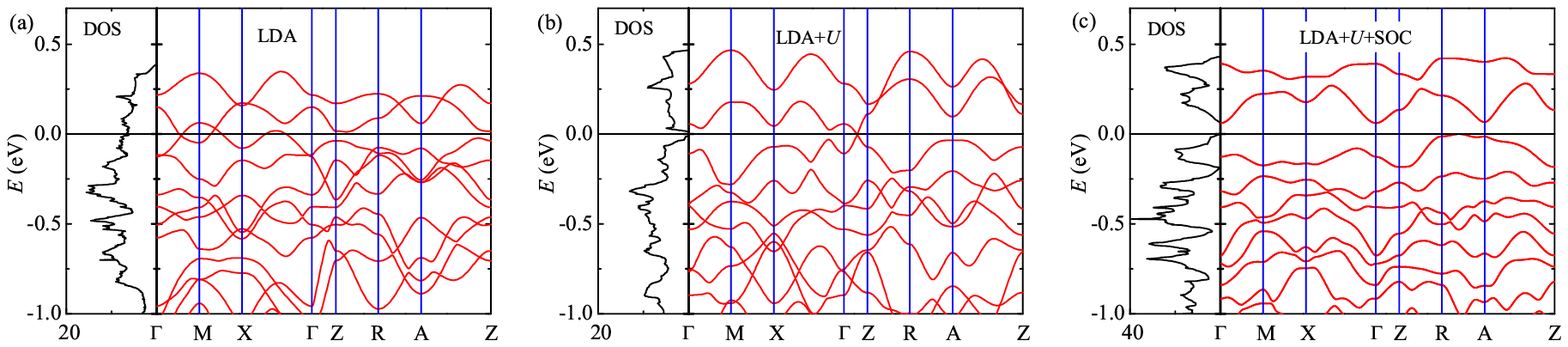}
\caption{(Color online)\label{Fig.5} The calculated density of state (DOS) and band structure of Li$_2$RhO$_3$, based on (a), LDA; (b), LDA+$U$ ($U=$3 eV); (c), LDA+$U$+SOC ($U=$3 eV). The calculations were performed with stripy-AFM structure.}
\end{figure*}

\begin{table}
\tabcolsep 0pt \caption{\label{table1} Rietveld refinement of Li$_2$RhO$_3$. Calculation based on space group C2/$m$ (No. 12). The derived $a=$5.1212(3)\AA,
$b=$8.8469(4)\AA, $c=$5.1015(3)\AA, $\alpha$=$\gamma=$90\textordmasculine, $\beta=$109.641(3)\textordmasculine. The quality factors of this refinement are $R_{wp}=$15.40\%, $R_{p}$=11.89\%, $S=$1.79 for \textbf{S1}, and $R_{wp}=$14.60\%, $R_{p}$=11.05\%, $S=$1.71 for \textbf{S2}.} \vspace*{-12pt}
\begin{center}
\def\temptablewidth{1.0\columnwidth}
{\rule{\temptablewidth}{1pt}}
\begin{tabular*}{\temptablewidth}{@{\extracolsep{\fill}}ccccccc}
Atom~                          & Site~  & $x$~   &$y$~   &$z$~    &occ.-\textbf{S1} &occ.-\textbf{S2} \\
\hline
Li(1)                          &  4g    & 0      & 0.333 & 0      & 0.185(5)        & 0.139(6)        \\
Li(2)                          &  2a    & 0      & 0     & 0      & 0.704(7)        & 0.721(5)        \\
Li(3)                          &  4h    & 0      & 0.833 & 0.5    &  1              &  1             \\
Li(4)                          &  2d    & 0      & 0.5   & 0.5    &  1              &  1             \\
Rh(1)                          &  4g    & 0      & 0.333 & 0      & 0.815(5)        & 0.861(6)        \\
Rh(2)                          &  2a    & 0      & 0     & 0      & 0.296(7)        & 0.279(5)        \\
O(1)                           &  8j    &0.266(1)& 0.333 &0.754(1)&  1              &  1             \\
O(2)                           &  4i    &0.266(1)& 0     &0.754(1)&  1              &  1             \\
\end{tabular*}
{\rule{\temptablewidth}{1pt}}
\end{center}
\end{table}

The electrical resistivity of \textbf{S1} is shown in Fig.~\ref{Fig.2}, from which an insulating $\rho$ vs. $T$ dependence is clearly seen. At 300 K, the magnitude of resistivity is 0.48 $\Omega$ cm, about two orders smaller than that of Na$_2$IrO$_3$\cite{Gegenwart-Na2IrO3}. In the inset of Fig.~\ref{Fig.2}, we show the $\rho$ vs. $1/T$ in the semi-logarithm plot, and we found that the $\rho(T)$ curve does not well follow the Arrhenius law for thermally-activated hopping, viz. $\rho(T)\propto\exp(\Delta/T)$, but is better fit to $\rho(T)\propto\exp[(D/T)^{1/4}]$ which is known as the three dimensional variable range hopping (3D-VRH)\cite{VRH} stemming from the random potential scattering contributed by large numbers of defects or disorders. Similar phenomenon was also observed in Na$_2$IrO$_3$\cite{Gegenwart-Na2IrO3}, and we attributed this to the antisite defect or disorder of Li and Rh, as is mentioned in Table~\ref{table1}. A rough estimate of activating energy $\Delta\sim$78 meV will be derived from the data 200-300 K. The measurement on \textbf{S2} leads to a similar result. According to a previous work performed by Todorova et al\cite{VTodorova-Li2RhO3}, the $\rho(T)$ curve obeys the Arrhenius law for the temperature region 300-500 K, and the derived energy gap is $\Delta=$ 80 meV. This magnitude of $\Delta$ is close to our result, and is much smaller than that of
Na$_2$IrO$_3$ (340 meV)\cite{Gegenwart-Na2IrO3MI}. All these confirm that Li$_2$RhO$_3$ is a narrow gap insulator.

The main frame of Fig.~\ref{Fig.3}(a) displays temperature dependence of dc magnetic susceptibility of \textbf{S1}. For the temperatures above 50 K, $\chi(T)$ obeys Curie-Weiss's law [see in the inset of Fig.~\ref{Fig.3}(a)], and can be well fitted to $\chi(T)=C/(T-\theta_W)$, where $\theta_W$ is the Weiss temperature. The fitting leads to $\theta_W=$-50 K. The negative $\theta_W$ indicates the dominant AFM coupling between Rh$^{4+}$ moments. The fitting also derives the effective moment $\mu_{eff}=$2.03 $\mu_B$ ($\mu_B$ is Bohr's magnon). This magnitude of $\mu_{eff}$ is close to but relatively larger than the ideal value for low-spin state of Rh$^{4+}$ (1.73 $\mu_B$) in the case of $J=S=$1/2 and Land\'{e} factor $g=$2, manifesting incompletely quenched orbital contribution. For the low temperature region, a sharp peak centered at 5.6 K is clearly seen in the curve for $\mu_0 H=$ 0.01 T. We also observed a discrepancy between ZFC and FC modes below this peak temperature. With increasing magnetic field, this peak looses sharpness and becomes rounded, and meanwhile, the discrepancy between ZFC and FC shrinks. This field dependent $\chi(T)$ was confirmed by the isothermal magnetization measurement shown in the inset of Fig.~\ref{Fig.3}(b). A tiny hysteresis loop with a remanent magnetization 1.4$\times$10$^{-3}\mu_B$/Rh and a coercive field 0.6 kOe is evident at 2 K. It should be pointed out that these observations were reproducible in different batches of samples [e.g., the result of \textbf{S2} is shown in Fig.~\ref{Fig.3}(b)],
although the peak position may vary slightly in the range of 5-7 K.

All these phenomena are hard to be understood by a simple AFM or ferromagnetic (FM) transition, but remind us of the spin-glass transition. We therefore performed the ac magnetic susceptibility of Li$_2$RhO$_3$, and the results are displayed in Fig.~\ref{Fig.3}(c). Indeed, a peak in the imaginary part of ac susceptibility $\chi''$ is evidently seen, strongly demonstrating the dissipative process. In addition, the real part of ac susceptibility $\chi'$ shows a peak at the freezing temperature ($T_f$) which shifts towards a higher temperature with increasing frequency ($f$). Such frequency dependent dissipative process is a fingerprint of a spin-glass transition. The frequency dependence of $T_f$ can be described by the conventional "critical slowing down" of the spin dynamics\cite{BY-slowing,Mydosh-SG,Gunnarsson}:
\begin{equation}
\tau(T_f)=\tau_0(T_f/T_g-1)^{-z\nu},
\label{Eq1}
\end{equation}
where $\tau=1/f$, $T_g$ is the characteristic temperature of spin-glass transition for $f\rightarrow$0, $z\nu$ is a dynamical exponent, while $\tau_0$ characterizes the intrinsic relaxation time of spin dynamics. We show this agreement by plotting $\log(f)$ vs. $\log[(T_f/T_g)-1]$ in the inset of Fig.~\ref{Fig.3}(c). The results of \textbf{S1} and \textbf{S2} can be well scaled into a same curve in this plot, with the parameters of this scaling law $z\nu=$8.2(3), $\tau_0\sim$10$^{-11.9(2)}$ s, while the critical temperatures for the two samples being $T_{g1}=$5.30 K and $T_{g2}=$5.97 K, respectively. The value of $z\nu$ is in good agreement with the theoretical prediction (7.0-8.0) for an Ising spin-glass system\cite{Kirkpatrick-zv,Campbell-zv}. The derived $\tau_0$ is in close approximation to that of a single atomic spin-glass system which usually possesses a $\tau_0$ in the order of 10$^{-13}$ s, implying that the observed spin-glass behavior is likely to arise from the frustrated single Rh$^{4+}$ ions in the honeycomb lattice, rather than from magnetic domains or clusters (for which $\tau_0$ can be as large as 10$^{-4}$ s\cite{Bera-SrNiVO_Ca}).

We now turn to the specific heat of Li$_2$RhO$_3$, as is shown in Fig.~\ref{Fig.4}. The measurement was carried out on the sample \textbf{S1}. An anomaly is clearly seen at around 7 K. Such anomaly differs from a $\lambda$-shaped specific heat jump usually seen in a second order phase transition, indicating the absence of long-range magnetic ordering and is consistent with the spin-glassy feature\cite{Fischer-SG}. Under magnetic field, this anomaly is suppressed and a Brillouin-like polarization trend is observable. Such evolution of specific heat under field provides further evidence to the competition between Zeeman energy and spin-glassy ordering, which is also depicted by the broadened spin-freezing peaks in $\chi(T)$ [Fig.~\ref{Fig.3}(a-b)]. We should emphasize that no long range magnetic ordering can be captured down to 0.5 K in Li$_2$RhO$_3$ by specific heat measurement. For comparison, the specific heat of its non-magnetic reference Li$_2$SnO$_3$ was also measured. We fit the specific heat of Li$_2$SnO$_3$ to the formula $C_{Sn}/T=\gamma_0^{Sn}+\beta^{Sn} T^2$, and the derived Sommerfeld coefficient is $\gamma_0^{Sn}=$0.18 mJ/(mol$\cdot$K$^2$). Such low $\gamma_0^{Sn}$ of Li$_2$SnO$_3$ signifies the highly insulating electronic property. The slope of this fit results in the Debye temperature $\Theta_D^{Sn}=$ 418 K. We calculated the lattice contribution to specific heat in Li$_2$RhO$_3$ by correcting $C_{Sn}$ to the molar mass\cite{MBouvier-SH}, and the estimated Debye temperature of Li$_2$RhO$_3$ is $\Theta_D^{Rh}=$ 444 K. The magnetic specific heat in Li$_2$RhO$_3$ is obtained by subtracting the lattice contribution from the total specific heat, and the result $C_{m}/T$ is displayed in the upper inset of Fig.~\ref{Fig.4}. The short range magnetic ordering above $T_g$ is further represented by the noticeable broad tail in $C_{m}/T$. We then calculated the magnetic entropy gain $S_m(T)$ by integrating $C_{m}/T$ over $T$. We found that $S_m$ reaches only 17\% of $R\ln 2$ at 20 K, and keeps increasing even for $T$ up to 50 K while no evident plateau can be seen (see the upper inset of Fig.~\ref{Fig.4}). There are two sources of entropy gain loss. Besides the short range magnetic ordering mentioned above, more magnetic entropy [$\sim$70\% of $R\ln2$, judging from $S_m(50 K)$] should be compensated by the residual magnetic entropy stemming from the quantum magnetic randomness that persists even at zero temperature, consistent with the spin-glass scenario.

To well understand these experimental results, we performed first-principles calculation\cite{CaoC-Li2RhO3}. For the $4d$ transition metal element, both electronic correlation and SOC should be taken into account. The calculated density of state (DOS) and band structure are shown in Fig.~\ref{Fig.5}. We started with the local density approximation (LDA), from which we derived a metallic electronic state [Fig.~\ref{Fig.5}(a)] as Rh$^{4+}$ has the half filled ionic configuration of $4d^5$. The application of Coulomb repulsion $U=$3 eV \cite{Dudarev} greatly reduces the DOS at Fermi level, however, there are still two bands crossing the Fermi level [Fig.~\ref{Fig.5}(b)] and forming a semi-metal-like band structure. We should point out that such semi-metal-like band structure is robust to Coulomb repulsion and will persist even under $U>$4 eV (data not shown). SOC was then employed, and the combination of $U$ and SOC successfully eliminates the band crossing at Fermi level and thus opens a gap $\Delta\sim$65 meV in the DOS spectrum, as shown in Fig.~\ref{Fig.5}(c). This magnitude of the energy gap is close to the thermal activating gap (78 meV) derived experimentally. The calculated SOC splitting is $\sim$10 meV, much smaller than that of Na$_2$IrO$_3$\cite{Gegenwart-Na2IrO3MI}, and the orbital- and spin- moment contributions are respectively $\langle L\rangle=$0.21 $\mu_B$/Rh and $\langle S\rangle$=0.19 $\mu_B$/Rh\cite{Wan-IridatesTI}. We should also point out that merely SOC can not open a gap at Fermi level, either (data not shown). Therefore, Li$_2$RhO$_3$ is suggestively to be a relativistic Mott insulator driven by both electronic correlation and SOC.

The calculation also helps us understand the spin-glassy feature of Li$_2$RhO$_3$. The study of such SOC mediated honeycomb lattice turns to the Heisenberg-Kitaev model\cite{Kitaev,Chaloupka-A2IrO3HK,Jiang-Na213phase,Liu-Na213Mag,Singh-A2IrO3HK}, in which FM ordering, AFM ordering with N\'{e}el-/ stripy-/ zigzag-type, or spin-liquid state emerges depending on the particular anisotropic magnetic couplings. In the case of Li$_2$RhO$_3$, the calculation based on LDA+$U$ prefers a FM ground state, but is challenged by several other magnetic configurations. When SOC is turned on, the ground state switches to stripy- or zigzag-type AFM configuration, and still many other magnetic configurations are comparable in energy\cite{CaoC-Li2RhO3}. In this situation, perturbations such as disorders or defects (see Table \ref{table1}) are likely to change the magnetic ground state. It is the fact that Li$_2$RhO$_3$ embeds in a regime close to the multi-phase boundary that results in the spin-glass nature. In addition, according to Choi et al's result of inelastic neutron scattering experiment on single crystalline Na$_2$IrO$_3$, there is some proportion of stacking fault of well-ordered honeycomb layers along $c$ axis\cite{Choi-Na2IrO3Str,Ye-Na2IrO3}, which is hardly resolvable by powder XRD pattern\cite{Gegenwart-Na2IrO3}. Such stacking fault might also appear in Li$_2$RhO$_3$ and account for the spin-glass ordering. To clarify the magnetism of Li$_2$RhO$_3$, single crystals are highly needed. With moderate strength of electronic correlation and SOC, our result sheds new light to the research of Heisenberg-Kitaev model in realistic materials and calls for more investigations in the future.


To summarize, we systematically studied the electronic and magnetic properties of Li$_2$RhO$_3$ on poly-crystalline samples. Our experiment confirms that Li$_2$RhO$_3$ is a spin-glassy insulator with a narrow gap $\Delta\sim$78 meV. This picture is supported by first-principles calculation which verifies the combination of electronic correlation and SOC. The calculation also points to many nearly degenerated magnetic configurations, which possibly illustrates the spin-glass behavior. Our result provides a unique case for the studies of Heisenberg-Kitaev model in realistic materials with moderate electronic correlation and SOC.


Y. Luo thanks Hui Xing and Ying Liu for helpful discussions. This work was supported by the National Basic Research Program of China
(Grant Nos. 2011CBA00103, 2012CB927404 and 2010CB923003), the National Science Foundation of China, and the Fundamental Research Funds for the Central Universities of China.


\begin{thebibliography}{10}

\bibitem{Bednorz-La2CuO4}
J. G. Bednorz, and K. A. M\"{u}ller, Z. Physik \textbf{B64}, 189 (1986).
\bibitem{Helmolt-LaMnO3}
R. von Helmolt, J. Wecker, B. Holzapfel, L. Schultz, and K. Samwer, Phys. Rev. Lett. \textbf{71}, 2331 (1993).
\bibitem{Maeno-Sr2RuO4}
Y. Maeno, H. Hashimoto, K. Yoshida, S. Nishizaki, T. Fujita, J. G. Bednorz, and F. Lichtenberg, Nature \textbf{372}, 532 (1994).
\bibitem{Nelson-Sr2RuO4}
K. D. Nelson, Z. Q. Mao, Y. Maeno, and Y. Liu, Science \textbf{306}, 1151 (2004).
\bibitem{Takada-NaxCoO2}
K. Takada, H. Sakurai, E. T. Muromachi, F. Izumi, R. A. Dilanian, and T. Sasaki, Nature \textbf{422}, 53 (2003).
\bibitem{Perry-Sr3Ru2O7}
R. S. Perry, L. M. Galvin, S. A. Grigera, L. Capogna, A. J. Schofield, A. P. Mackenzie, M. Chiao, S. R. Julian, S. I. Ikeda, S. Nakatsuji, Y. Maeno, and C. Pfleiderer, Phys. Rev. Lett. \textbf{86}, 2661 (2001).
\bibitem{Yanagishima-R2Ir2O7}
D. Yanagishima, and Y. Maeno, J. Phys. Soc. Jpn. \textbf{70}, 2880 (2001).
\bibitem{Erickson-BaNaOsO}
A. S. Erickson, S. Misra, G. J. Miller, R. R. Gupta, Z. Schlesinger, W. A. Harrison, J. M. Kim, and I. R. Fisher, Phys. Rev. Lett. \textbf{99}, 016404 (2007).
\bibitem{Gegenwart-Na2IrO3}
Y. Singh, and P. Gegenwart, Phys. Rev. B \textbf{82}, 064412 (2010).
\bibitem{Kim-Sr2IrO4}
B. J. Kim, Hosub Jin, S. J. Moon, J. Y. Kim, B. G. Park, C. S. Leem, J. Yu, T. W. Noh, C. Kim, S. J. Oh, J. H. Park, V. Durairaj, G. Cao, and E. Rotenberg, Phys. Rev. Lett. \textbf{101}, 076402 (2008).
\bibitem{Perry-Sr2RhO4}
R. S. Perry, F. Baumberger, L. Balicas, N. Kikugawa, N. J. C. Ingle, A. Rost, J. F. Mercure, Y. Maeno, Z. X. Shen, and A. P. Mackenzie, New J. Phys. \textbf{8}, 175 (2006).
\bibitem{Jansen-Li2MnO3}
M. Jansen, R. Hoppe, Z. Anorg. Allg. Chem. \textbf{397}, 279 (1973).
\bibitem{Lang-Li2SnO3}
G. Lang, Z. Anorg. Allg. Chem. \textbf{348}, 246 (1966).
\bibitem{VTodorova-Li2RhO3}
V. Todorova, and M. Jansen, Z. Anorg. Allg. Chem. \textbf{637}, 37 (2011).
\bibitem{Kobayashi-Li2RuO3}
H. Kobayashi, R. Kanno , Y. Kawamoto, M. Tabuchi, O. Nakamura, and M. Takano, Solid State Ionics \textbf{82}, 25 (1995).
\bibitem{Kobayashi-Li2IrO3}
H. Kobayashi, M. Tabuchi, M. Shikano, H. Kageyamaa, and R. Kannob, J. Mater. Chem. \textbf{13}, 957(2003).
\bibitem{Malley-Li2IrO3}
M. J. O$'$Malley, H Verweij, and P. M. Woodward, J. Solid State Chem. \textbf{181}, 1803 (2008).
\bibitem{Izumi-Rietan_RF}
F. Izumi, and K. Momma, Solid State Phenom. \textbf{130}, 15 (2007).
\bibitem{VRH}
P. V. E. McLintock, D. J. Meredith, and J. K. Wigmore, {\it Matter at low temperatures}, (Blcakie, Glasgow, 1984), pp. 82.
\bibitem{Gegenwart-Na2IrO3MI}
R. Comin, G. Levy, B. Ludbrook, Z. H. Zhu, C. N. Veenstra, J. A. Rosen, Yogesh Singh, P. Gegenwart, D. Stricker, J. N. Hancock, D. van der Marel, I. S. Elfimov, and A. Damascelli, Phys. Rev. Lett. \textbf{109}, 266406 (2012).
\bibitem{BY-slowing}
K. Binder, and A. P. Young, Phys. Rev. B \textbf{29}, 2864 (1984).
\bibitem{Mydosh-SG}
J. A. Mydosh, {\it Spin Glasses: An Experiment Introduction} (Taylor \& Francis, London, 1993).
\bibitem{Gunnarsson}
K. Gunnarsson, P. Svedlindh, P. Nordblad, L. Lundgren, H. Aruga, and A. Ito, Phys.Rev. Lett. \textbf{61}, 754 (1988).
\bibitem{Kirkpatrick-zv}
S. Kirkpatrick, and D. Sherrington, Phys. Rev. B \textbf{17}, 4384 (1978).
\bibitem{Campbell-zv}
I. A. Campbell, Phys. Rev. B \textbf{33}, 3587 (1986).
\bibitem{Bera-SrNiVO_Ca}
A. K. Bera, and S. M. Yusuf, Phys. Rev. B \textbf{86}, 024408 (2012).
\bibitem{Fischer-SG}
K. H. Fischer, and J. A. Hertz, {\it Spin Glasses} (Cambridge University Press, Cambridge, 1991) p.301.
\bibitem{MBouvier-SH}
M. Bouvier, P. Lethuillier, and D. Schmitt, Phys. Rev. B \textbf{43}, 13137 (1991).
\bibitem{CaoC-Li2RhO3}
The calculations were performed with plane-wave basis projected-augmented wave code VASP, with the PBE flavor of GGA chosen to be the exchange-correlation functional. Total energy cutoff was set to be 540 eV and 8$\times$5$\times$8 $\Gamma$-centered $K$-mesh was used. For more details of the calculation, please refer to C. Cao, Y. Luo, Z. Xu, and J. Dai, arXiv: 1303.4675 (2013).
\bibitem{Dudarev}
S. L. Dudarev, G. A. Botton, S. Y. Savrasov, C. J. Humphreys and A. P. Sutton, Phys. Rev. B \textbf{57}, 1505 (1998).
\bibitem{Wan-IridatesTI}
X. Wan, A. M. Turner, A. Vishwanath, and S. Y. Savrasov, Phys. Rev. B, \textbf{83}, 205101 (2011).
\bibitem{Kitaev}
A. Kitaev, Ann. Phys. (N.Y.) \textbf{321}, 2 (2006).
\bibitem{Chaloupka-A2IrO3HK}
J. Chaloupka, G. Jackeli, and G. Khaliullin, Phys. Rev. Lett. \textbf{105}, 027204 (2010).
\bibitem{Jiang-Na213phase}
H. C. Jiang, Z. C. Gu, X. L. Qi, and S. Trebst, Phys. Rev. B \textbf{83}, 245104 (2011).
\bibitem{Liu-Na213Mag}
X. Liu, T. Berlijn, W. G. Yin, W. Ku, A. Tsvelik, Y. J. Kim, H. Gretarsson, Y. Singh, P. Gegenwart, and J. P. Hill, Phys. Rev. B \textbf{83}, 220403(R) (2011).
\bibitem{Singh-A2IrO3HK}
Y. Singh, S. Manni, J. Reuther, T. Berlijn, R. Thomale, W. Ku, S. Trebst, and P. Gegenwart, Phys. Rev. Lett. \textbf{108}, 127203 (2012).
\bibitem{Choi-Na2IrO3Str}
S. K. Choi, R. Coldea, A. N. Kolmogorov, T. Lancaster, I. I. Mazin, S. J. Blundell, P. G. Radaelli, Y. Singh, P. Gegenwart, K. R. Choi, S. W. Cheong, P. J. Baker, C. Stock, and J. Taylor, Phys. Rev. Lett. \textbf{108}, 127204 (2012).
\bibitem{Ye-Na2IrO3}
F. Ye, S. Chi, H. Cao, B. C. Chakoumakos, J. A. Fernandez-Baca, R, Custelcean, T. F. Qi, O. B. Korneta, and G. Cao, Phys. Rev. B \textbf{85}, 180403(R) (2012).

\end{thebibliography}
\end{document}